\documentclass[a4paper,11pt]{article}
\usepackage[utf8]{inputenc}

\usepackage[margin=2.5cm]{geometry}

\usepackage[american]{babel}
\usepackage{csquotes}

\usepackage[utf8]{inputenc}

\usepackage[ttscale=.875]{libertine}
\usepackage[scaled=0.96]{zi4}

\usepackage[backend=biber,
            style=apa,
            maxcitenames=3,
            maxbibnames=99,
            apamaxprtauth=99,
            natbib=true]{biblatex}
\DeclareLanguageMapping{american}{american-apa}
\bibliography{draft.bib}

\usepackage{graphicx}

\usepackage{listings}
\usepackage[framemethod=tikz]{mdframed}
\makeatletter
\def\mdf@@codeheading{Code Listings}
\define@key{mdf}{title}{\def\mdf@@codeheading{#1}}
\mdfdefinestyle{lstlisting}{%
  backgroundcolor=black!2.5,
  innertopmargin=2pt,
  middlelinewidth=0.75pt,
  outerlinewidth=9pt,
  outerlinecolor=white,
  innerleftmargin=10pt,
  innerrightmargin=10pt,
  leftmargin=0pt,
  rightmargin=0pt,
  rightline=false,
  leftline=false,
  bottomline=false,
  skipabove=\topskip,
  skipbelow=\topskip,
  roundcorner=0pt,
  singleextra={
    \node[text=black, 
          anchor=south west, yshift=-0.25pt, xshift=5pt,
          font=\footnotesize] at (O|-P) {\csname mdf@@codeheading\endcsname};},
  firstextra={
    \node[text=black, 
          anchor=south west, yshift=-0.5pt, xshift=5pt,
          font=\footnotesize] at (O|-P) {\csname mdf@@codeheading\endcsname};}
}

\lstnewenvironment{code}[2][]{%
  \lstset{#1}%
  \mdframed[style=lstlisting,title={#2}]%
}{\endmdframed}

\usepackage{xcolor}
\definecolor{blendedblue}{rgb}{0.2, 0.2, 0.6}
\definecolor{blendedred}{rgb}{0.8, 0.2, 0.2}
\usepackage[bookmarks=true,
            breaklinks=true,
            pdfborder={0 0 0},
            citecolor=blendedblue,
            colorlinks=true,
            linkcolor=blendedblue,
            urlcolor=blendedblue,
            citecolor=blendedblue,
            linktocpage=false,
            hyperindex=true,
            linkbordercolor=white]{hyperref}
\usepackage{hyperref}
\hypersetup{colorlinks=true}

\title{Re-run, Repeat, Reproduce, Reuse, Replicate:\\Transforming Code into Scientific Contributions}
\author{
  \textbf{Fabien C. Y. Benureau}$^{1,2,3,*}$,
  \textbf{Nicolas P. Rougier}$^{1,2,3}$\\
  {\small \href{https://orcid.org/0000-0003-4083-4512}{0000-0003-4083-4512},
  \href{https://orcid.org/0000-0002-6972-589X}{0000-0002-6972-589X}}\\
  \begin{minipage}{0.8\textwidth}
    \begin{center}
      \vspace{2mm}
      \scriptsize
      $^{1}$INRIA Bordeaux Sud-Ouest, Talence, France\\
      $^{2}$ Institut des Maladies Neurodégénératives, Université de Bordeaux, CNRS UMR 5293, Bordeaux, France\\
      $^{3}$ LaBRI, Université de Bordeaux, Bordeaux INP, CNRS UMR 5800, Talence, France\\
      \vspace{2mm}
      $^{*}$Corresponding author:
      \href{mailto:fabien.benureau@gmail.com}{fabien.benureau@gmail.com}
    \end{center}
  \end{minipage}
}
\date{}

\begin{document}
\maketitle
\section*{Introduction (R$^{\mathbf 0}$)}

Replicability is a cornerstone of science.  If an experimental result cannot be
re-obtained by an independent party, it merely becomes, at best, an observation
that may inspire future research \citep{Mesirov:2010,osc:2015}. Replication
issues have received increased attention in recent years, with a particular focus on medicine and psychology \citep{Iqbal:2016}.  One could think
that computational research would mostly be shielded from such issues, since a
computer program describes precisely what it does and is easily
disseminated to other researchers without alteration.

But precisely because it is easy to believe that if a program runs once and
gives the expected results it will do so forever, crucial steps to transform
working code into meaningful scientific contributions are rarely undertaken
\citep{Collberg:2016,Sandve:2013,Schwab:2000}. Computational research is plagued by replication
problems, in part, because it seems impervious to them. Contrary to production software who provide a service geared towards a practical outcome , the motivation behind scientific code is to test an hypothesis. While in some instance production software and scientific code are indistinguishable, the reason why they were created is different, and, therefore, so are the criteria to evaluate their success. A program can
fail as a scientific contribution in many different ways for many different
reasons. Borrowing the terms coined by \citeauthor{Goble:2016}
\citep{Goble:2016}, for a program to contribute to science, it should be
re-runnable (R$^1$), repeatable (R$^2$), reproducible (R$^3$),
reusable (R$^4$) and replicable (R$^5$). Let us illustrate this with a small
example, a random walk \citep{Hughes:1995} written in Python:\\

\noindent \begin{minipage}[c]{\linewidth}
\begin{code}{\parbox{.8\textwidth}{\textbf{\textsc{Listing 1:}} Random walk (R$^0$)}\parbox{.161\textwidth}{\hfill \href{https://raw.githubusercontent.com/rougier/random-walk/frontiers/random-walk-R0.py}{raw code}, \href{https://doi.org/10.5281/zenodo.848217}{archive}}}
import random

x = 0
for i in xrange(10):
    step = random.choice([-1,+1])
    x += step
    print x,
\end{code}
\end{minipage}

In the code above, the {\tt \href{https://docs.python.org/3.6/library/random.html#random.choice}{random.choice}} function randomly returns either +1 or -1. The instruction \enquote{\tt for i in xrange(10):}  executes the next three indented lines ten times. Executed, this program would display:
\begin{code}{Output}
-1, 0, -1, 0, -1, 0, -1, 0, 1, 2 # with the steps being -1, +1, -1, +1, -1, +1, -1, +1, +1, +1
\end{code}

What could go wrong with such a simple program?\\
\vfill
Well...
\vfill

\clearpage
\section*{Re-runnable (R$^{\mathbf 1}$)}

Have you ever tried to re-run a program you wrote some years ago? It can often be frustratingly hard. Part of the problem is that technology is evolving at a fast pace and you cannot know in advance how the system, the software and the libraries your program depends on will evolve. Since you wrote the code, you may have reinstalled or upgraded  your operating system. The compiler, interpreter or set of libraries installed may have been replaced with newer versions. You may find yourself battling with arcane issues of library compatibility---thoroughly orthogonal to your immediate research  goals---to execute again a code \emph{that worked perfectly before}. To be clear, it is impossible to write future-proof code, and the best efforts can be stymied by the smallest change in one of the dependencies. At the same time, modernizing an unmaintained ten-year-old code can reveal itself to be an arduous and expensive undertaking---and precarious, since each change risks affecting the semantics of the program. Rather than trying to predict the future or painstakingly dusting off old code, an often more straightforward solution is to recreate the old execution environment\footnote{To be clear, and although virtual machines are often a great help here, this is not always possible. It is, however, \emph{always} more difficult when the original execution environment is unknown.}. For this to happen however, the dependencies in terms of systems, software and libraries must be made clear enough.\\

A \emph{re-runnable} code is one that can be run again when needed, and in particular more than the one time that was needed to produce the results. It is important to notice that the re-runnability of a code is not an intrinsic property. Rather, it depends on the context, and becomes increasingly difficult as the code ages. Therefore, to be and remain re-runnable on other researchers' computers, a re-runnable code should describe---with enough details to be recreated---an execution environment in which it is executable. As shown by \citep{Collberg:2016}, this is far from being either obvious or easy.\\

\noindent \begin{minipage}[c]{\linewidth}
\begin{code}{\parbox{.8\textwidth}{\textbf{\textsc{Listing 2:}} Re-runnable random walk (R$^1$)}\parbox{.161\textwidth}{\hfill \href{https://raw.githubusercontent.com/rougier/random-walk/frontiers/random-walk-R1.py}{raw code}, \href{https://doi.org/10.5281/zenodo.848217}{archive}}}
# Tested with Python 3
import random

x =  0
walk = []
for i in range(10):
    step = random.choice([-1,+1])
    x += step
    walk.append(x)

print(walk)
\end{code}
\end{minipage}\\

In our case, the R$^0$ version of our tiny walker seems to imply that any version of Python would be fine. This not the case: it uses the print {\em instruction} and the {\tt xrange} operator, both specific to Python 2. The print {\em instruction}, available in Python 2 (a version still widely used; support is scheduled to stop in 2020), has been deprecated in Python 3 (first released in 2008, almost a decade ago) in favor or a  print {\em function}, while the {\tt xrange} operator has been replaced by the {\tt range} operator in Python 3. In order to try to future-proof the code a bit, we might as well target Python 3, as is done in the R$^1$ version. Incidentally, it remains compatible with Python 2. But whichever version is chosen, the crucial step here is to document it.

\section*{Repeatable (R$^{\mathbf 2}$)}

The code is running and producing the expected results. The next step is to make sure that you can produce the same output over successive runs of your program. In other words, the next step is to make your program deterministic, producing {\em repeatable} output.  Repeatability is valuable. If a run of the program produces a particularly puzzling result, repeatability allows you to scrutinize any step of the execution of the program by re-running it again with extraneous prints, or inside a debugger. Repeatability is also useful to prove that the program did indeed produce the published results. Repeatability is not always possible or easy \citep{Diethelm:2012, Courtes:2015}. But for sequential and deterministically parallels programs \citep{Hines:2008, Collange:2015} not depending on analog inputs, it often comes down to controlling the initialization of the pseudo-random number generators (RNG).\\

For our program, that means setting the seed of the {\tt random} module. We may also want to save the output of the program to a file, so that we can easily verify that consecutive runs do produce the same output: eyeballing differences is unreliable and  time-consuming, and therefore won't be done systematically.\\

\noindent \begin{minipage}[c]{\linewidth}
\begin{code}{\parbox{.8\textwidth}{\textbf{\textsc{Listing 3:}} Re-runnable, repeatable random walk (R$^2$)}\parbox{.161\textwidth}{\hfill \href{https://raw.githubusercontent.com/rougier/random-walk/frontiers/random-walk-R2.py}{raw code}, \href{https://doi.org/10.5281/zenodo.848217}{archive}}}
# Tested with Python 3
import random

random.seed(1) # RNG initialization

x =  0
walk = []
for i in range(10):
    step = random.choice([-1,+1])
    x += step
    walk.append(x)

print(walk)
# Saving output to disk
with open('results-R2.txt', 'w') as fd:
    fd.write(str(walk))
\end{code}
\end{minipage}\\

Setting seeds should be done carefully. Using 439 as a seed in the previous program would result in ten consecutive +1 steps\footnote{With CPython 3.3-3.6. See the next section for details.}, which---although a perfectly valid random walk---lend itself to a gross misinterpretation of the overall dynamics of the algorithm. Verifying that the qualitative aspects of the results and the conclusions that are made are not tied to a specific initialization of the pseudo-random generator is an integral part of any scientific undertaking in computational science;
this is usually done by repeating the simulations multiple times with different seeds.


\section*{Reproducible (R$^{\mathbf 3}$)}

The R$^2$ code seems fine enough, but it hides several problems that come to light when trying to {\em reproduce} results. A result is said to be \emph{reproducible} if another researcher can take the original code and input data, execute it, and re-obtain the same result \parencite{Peng:2006}. As explained by \citeauthor{Donoho:2009} \parencite{Donoho:2009}, scientific practice must expect that {\em errors are ubiquitous}, and therefore be robust to them. Ensuring reproducibility is a fundamental step toward this: it provides other researchers the means to verify that the code does indeed produce the published results, and to scrutinize the procedures it used to produce them. As demonstrated by \citeauthor{Mesnard:2016} \citep{Mesnard:2016}, reproducibility is hard.\\

For instance, the R$^2$ program will not produce the same results all the time. It will, because it is repeatable, produce the same results over repeated executions. But it will not necessarily do so over different execution environments. The cause is to be found in a change that occurred in the pseudo-random number generator between Python 3.2 and Python 3.3. Executed with Python 2.7 to 3.2, the code will produce the sequence -1, 0, 1, 0, -1, -2, -1, 0, -1, -2. But with Python 3.3 to 3.6, it will produce -1, -2, -1, -2, -1, 0, 1, 2, 1, 0. With future versions of the language, it may change still. For the R$^3$ version, we abandon the use of the {\tt \href{https://docs.python.org/3.6/library/random.html#random.choice}{random.choice}} function in favor of the {\tt \href{https://docs.python.org/3.6/library/random.html#random.uniform}{random.uniform}} function, whose behavior is consistent across versions 2.7 to 3.6 of Python.\\

Because any dependency of a program---to the most basic one, the language itself---can change its behavior from one version to another, executability (R$^1$) and determinism (R$^2$) are necessary but not sufficient for reproducibility. The exact execution environment used to produce the results must also be specified---rather than the broadest set of environments where the code can be effectively run. In other words, assertions such as \enquote{the results were obtained with CPython 3.6.1} are more valuable, in a scientific context, than \enquote{the program works with Python 3.x and above}. With the increasing complexity of computational stacks, retrieving and deciding what is pertinent (CPU architecture? operating system version? endianness?) might be non-trivial. A good rule of thumb is to include more information than necessary rather than not enough, and some rather than none.\\

Recording the execution environment is only the first step. The R$^2$ program uses a random seed but does not keep a trace of it except in the code. Should the code change after the production of the results, someone provided with the last version of the code will not be able to know which seed was used to produce the results, and would need to iterate through all possible random seeds, an impossible task in practice\footnote{Here, with $2^{10}$ possibilities for a 10-step random walk, a seed matching the any possible sequence could certainly be found. For instance, seed 11235813 matches the results of R$^0$ with Python 2.7. Such a search becomes intractable for a 100-step walk.}.\\

This is why result files should come alongside their context, i.e. an exhaustive list of the parameters used as well as a precise description of the execution environment, as the R$^3$ code does. The code itself is part of that context: the version of the code must be recorded. It is common for different results or different figures to have been generated by different versions of the code. Ideally, all results should originate from the same (and last) version of the code. But for long or expensive computations, this may not be feasible. In that case, the result files should contain the version of the code that was used to produce it. This information can be obtained from the version control software. This also allows, if some errors are found and corrected after some results have been obtained, to identify which ones should be recomputed. In R$^3$, the code records the git revision, and whether the repository holds uncommitted changes when the computation starts.\\

Published results should obviously come from version of the code where every change and every file has been committed. This includes pre-processing, post-processing and plotting code. Plotting code may seem mundane, but it is as vulnerable as any other piece of the code to bugs and errors.
When it comes to checking that the reproduced data match the one published in the article, however, figures can reveal themselves to be imprecise and cumbersome, and sometimes plain unusable. To avoid having to manually overlay pixelated plots, published figures should be accompanied by their underlying data (coordinates of the plotted points) in the supplementary data to allow straightforward numeric comparisons.\\

Another good practice is to make the code self-verifiable. In R$^3$, a short unit test is provided, that allows the code to verify its own reproducibility. Should this test fail, then there is little hope of reproducing the results. Of course, passing the test does not guarantee anything.\\

It is obvious that \emph{reproducibility implies availability}. As shown in \citep{Collberg:2016}, code is often unavailable, or only available upon request. While the latter may seem sufficient, changes in email address, changes in career, retirement, a busy inbox or poor archiving practices can make a code just as unreachable. Code \emph{and} input data \emph{and} result data should be available with the published article, as supplementary data, or through a DOI link to a scientific repository such as \href{https://figshare.com}{Figshare}, \href{https://zenodo.org}{Zenodo}\footnote{Online code repositories such as GitHub \emph{are not} scientific repositories, and may disappear, change name or change their access policy at any moment. Direct links to them are not perpetual, and, when used, they should always be supplemented by a DOI link to a scientific archive.} or a domain specific database, such as \href{https://senselab.med.yale.edu/modeldb/}{ModelDB} for computational neuroscience. The codes presented in this article are available in the GitHub repository \href{https://github.com/rougier/random-walk}{github.com/rougier/random-walk} and at \href{https://doi.org/10.5281/zenodo.848217}{doi.org/10.5281/zenodo.848217}.\\

\noindent \begin{minipage}[c]{\linewidth}
\begin{code}{\parbox{.8\textwidth}{\textbf{\textsc{Listing 4:}} Re-runnable, repeatable, reproducible random walk (R$^3$)}\parbox{.161\textwidth}{\hfill \href{https://raw.githubusercontent.com/rougier/random-walk/frontiers/random-walk-R3.py}{raw code}, \href{https://doi.org/10.5281/zenodo.848217}{archive}}}
# Copyright (c) 2017 Nicolas P. Rougier and Fabien C. Y. Benureau
# Release under the BSD 2-clause license
# Tested with CPython 3.6.2 / macOS 10.12.6 / 64 bits architecture
import sys, subprocess, datetime, random

def generate_walk():
    x = 0
    walk = []
    for i in range(10):
        if random.uniform(-1, +1) > 0:
            x += 1
        else:
            x -= 1
        walk.append(x)
    return walk

# If repository is dirty, don't run anything
if subprocess.call(('git', 'diff-index', '--quiet', 'HEAD')):
    print('Repository is dirty, please commit first')
    sys.exit(1)

# Get git hash if any
revision = subprocess.check_output(('git', 'rev-parse', 'HEAD'))

# Unit test
random.seed(42)
assert generate_walk() == [1, 0, -1, -2, -1, 0, 1, 0, -1, -2]

# Random walk for 10 steps
seed = 1
random.seed(seed)
walk = generate_walk()

# Display & save results
print(walk)
results = {'data'       : walk,
            'seed'      : seed,
            'timestamp': str(datetime.datetime.utcnow()),
            'revision' : revision,
            'system'    : sys.version}
with open('results-R3.txt', 'w') as fd:
    fd.write(str(results))
\end{code}
\end{minipage}\\

To recap, reproducibility implies re-runnability and repeatability and availability, yet imposes additional conditions. Dependencies and platforms must be described as precisely and as specifically as possible. Parameters values and inputs should accompany the result files. The data and scripts behind the graphs must be published. Unit tests are a good way to embed self-diagnostics of reproducibility in the code. Reproducibility is hard, yet tremendously necessary.

\section*{Reusable (R$^{\mathbf 4}$)}

Making your program reusable means it can be easily used, and modified, by you and other people, inside and outside your lab. Ensuring your program is reusable is advantageous for a number of reasons.\\

For you, first. Because the you now and the you in two years are two different persons. Details on how to use the code, its limitations, its quirks, may be present to your mind now, but will probably escape you in six months \parencite{Donoho:2009}. Here, comments and documentation can make a significant difference. Source code reflects the results of the decisions that were made during its creation, but not the reasons behind those decisions. In science, where the method and its justification matter as much as the results, those reasons are precious knowledge. In that context, a comment on how a given parameter was chosen (optimization, experimental data, educated guess), why a library was chosen over another (conceptual or technical reasons?) is valuable information.\\


Reusability of course directly benefits other researchers from your team and outside of it. The easier it is to use your code, the lower the threshold is for other to study, modify and extend it. Scientists constantly face the constraint of time: if a model is available, documented, and can be installed, run and understood all in a few hours, it will be preferred over another that would require weeks to reach the same stage. A reproducible and reusable code offers a platform both \emph{verifiable} and easy-to-use, fostering the development of derivative works by other researchers on solid foundations. Those derivative works contribute to the impact of your original contribution.\\

Having more people examining and using your code also means that potential errors have a higher chance to be caught. If people start using your program, they will most likely report bugs or malfunctions they encounter. If you're lucky enough, they might even propose either bug fixes or improvements, hence improving the overall quality of your software. This process contributes to the long-term reproducibility to the extent people continue to use and maintain the program.\\

Despite all this, reusability is often overlooked, and it is not hard to see why. Scientists are rarely trained in software engineering, and reusability can represent an expensive endeavour if undertaken as an afterthought, for little tangible short-term benefits, for a codebase that might, after all, see only a single use. And, in fact, reusability is not as indispensable a requirement as re-runnability, repeatability and reproducibility. Yet, some simple measures can tremendously increase reusability, and at the same time strengthen reproducibility and re-runnability over the long-term.\\

Avoid hardcoded or magic numbers. Magic numbers are numbers present directly in the source code, that do not have a name and therefore can be difficult to interpret semantically. Hardcoded values are variables that cannot be changed through a function argument or a parameter configuration file. To be modified, they involve editing the code, which is cumbersome and error-prone. In the R$^3$ code, the seed and the number of steps are respectively hardcoded and magic.\\

Similarly, code behavior should not be changed by commenting/uncommenting code \citep{Wilson:2017}. Modification of the behavior of the code, required when different experiments examine slightly different conditions, should always be explicitly set through parameters accessible to the end-user. This improves reproducibility in two ways: it allows those conditions to be recorded as parameters in the result files, and it allows to define separate scripts to run or configuration files to load to produce each of the figures of the published paper. With documentation explaining which script or configuration file corresponds to which experiment, reproducing the different figures becomes straightforward.\\

\noindent \begin{minipage}[c]{\linewidth}
\begin{code}{\parbox{.8\textwidth}{\textbf{\textsc{Listing 5:}} Re-runnable, repeatable, reproducible, reusable random walk (R$^4$)}\parbox{.161\textwidth}{\hfill \href{https://raw.githubusercontent.com/rougier/random-walk/frontiers/random-walk-R4.py}{raw code}, \href{https://doi.org/10.5281/zenodo.848217}{archive}}}
# Copyright (c) 2017 Nicolas P. Rougier and Fabien C.Y. Benureau
# Release under the BSD 2-clause license
# Tested with CPython 3.6.2 / macOS 10.12.6 / 64 bits architecture
import sys, subprocess, datetime, random

def generate_walk(count, x0=0, step=1, seed=0):
    """ Random walk
        count: number of steps
        x0   : initial position (default 0)
        step : step size (default 1)
        seed : seed for the initialization of the random generator (default 0)
    """
    random.seed(seed)
    x = x0
    walk = []
    for i in range(count):
        if random.uniform(-1, +1) > 0:
            x += 1
        else:
            x -= 1
        walk.append(x)
    return walk

def generate_results(count, x0=0, step=1, seed=0):
    """Compute a walk and return it alongside its context"""
    # If repository is dirty, don't do anything
    if subprocess.call(('git', 'diff-index', '--quiet', 'HEAD')):
        print('Repository is dirty, please commit first')
        sys.exit(1)

    # Get git hash if any
    revision = subprocess.check_output(('git', 'rev-parse', 'HEAD'))

    # Compute results
    walk = generate_walk(count=count, x0=x0, step=step, seed=seed)
    return {'data'      : walk,
            'parameters': {'x0':x0, 'step':step, 'count':count, 'seed':seed},
            'timestamp' : str(datetime.datetime.utcnow()),
            'revision'  : revision,
            'system'    : sys.version}

if __name__ == '__main__':
    # Unit test checking reproducibility (will fail with Python<=3.2)
    assert generate_walk(10, 0, 1, 42) == [1, 0, -1, -2, -1, 0, 1, 0, -1, -2]

    # Simulation parameters
    count, x0, seed = 10, 0, 1
    results = generate_results(count, x0=x0, seed=seed)

    # Save & display results
    with open('results-R4.txt', 'w') as fd:
        fd.write(str(results))
    print(results['data'])
\end{code}
\end{minipage}\\

Documentation is one of the most potent tools for reusability. A proper documentation on how to install and run the software often makes the difference whether other researchers manage to use it or not. A comment describing what each function does, however evident, can avoid hours of head-scratching. Great code may need few comments. Scientists, however, are not always brilliant developers. Of course, bad, complicated code should be rewritten until is simple enough to explain itself. But realistically, this is not always going to be done: there is simply not enough incentive for it. There, a comment that explains the intentions and reasons behind a block of code can be tremendously useful.\\

Reusability is not a strict requirement for scientific code. But it has many benefits, and a few simple measures can foster it considerably. To complement the R$^4$ version provided here, we provide an example repository of a re-runnable, repeatable, reproducible and reusable random walk code. The repository is available on GitHub \href{https://github.com/benureau/r5}{github.com/benureau/r5} and here \href{https://doi.org/10.5281/zenodo.848284}{doi.org/10.5281/zenodo.848284}.

\section*{Replicable (R$^{\mathbf 5}$)}

Having made a software reusable offers an additional way to find errors, especially if your scientific contribution is popular. Unfortunately, this is not always effective, and some recent cases have shown that bugs can lurk in well-used open-source code, impacting the false positive rates of fMRI studies \citep{Eklund:2016}, or the encryption of communications over the Internet \citep{Durumeric:2014}. Let's be clear: the goal here is not to remove all bugs and mistakes from science. The goal is to have methods and practices in place that make possible for the inevitable errors that will be made to be caught and corrected by motivated investigators. This is why, as explained by Peng et al. \citep{Peng:2006}, {\em the replication of
  important findings by multiple independent investigators is fundamental to
  the accumulation of scientific evidence}.\\

\emph{Replicability} is the implicit
assumption that an article that does not provide the source code makes: that
the description it provides of the algorithms is sufficiently precise and
complete to re-obtain the results it presents. Here, replicating implies writing a new code matching the conceptual description of the article, in order to reobtain the same results. Replication affords robustness to the results because, should the original code contain an error, a different codebase creates the possibility that this error will not be repeated, in the same way that replicating an laboratory experiment in a different laboratory can ferret out subtle biases. While every published article should strive for replicability, it is seldom obtained. In fact, absent an explicit effort to make an algorithmic description replicable, there is little probability that it will be.\\

This is because most papers strive to communicate the main ideas behind their contribution in terms as simple and as clear as possible, so that the reader may be able to easily understand them and the results that are presented. Trying to ensure replicability in the main text adds a myriad of esoteric details that are not conceptually significant and clutter the explanations. Therefore, unless the writer dedicates an addendum or a section of the supplementary information for technical details specifically aimed at replicability, the information will not be there because there are incentives not to do so.\\

But even when those details are present, the best efforts may fall short because an oversight, a typo or a difference between what is evident for the writer and for the reader \citep{Mesnard:2016}. Minute changes in the numerical estimation of a common first-order differential equation can have significant impact \citep{Crook:2013}. Hence, a reproducible code plays an important role alongside its article: it is a objective catalog of all the implementation details.\\

A researcher seeking to replicate published results might first consider only the article. If she fails to replicate the results, she will consult the original code, and with it be able to pinpoint why her code and the code of the authors differ in behavior. Because a mistake on their part? Hers? Or a difference in a seemingly innocuous implementation detail? A fine analysis of why a particular algorithmic description is lacking or ambiguous or why a minor implementation decision is in fact crucial to obtain the published results is of great scientific value. Such an analysis can only be done with access to both the article and the code. With only the article, the researcher will often be unable to understand why she failed to replicate the results, and will naturally be inclined to only report replication successes.\\

Replicability, therefore, does not negate the necessity of reproducibility. In fact, it often relies on it. To illustrate this, let us consider what could be the textual description of the random walker (as it would be written in an article describing it):
\begin{quotation}
{\em The model uses the Mersene Twister generator initialized with the seed 1. At each iteration, a uniform number between -1 (included) and +1 (excluded) is drawn and the sign of the result is used for generating a positive or negative step.}
\end{quotation}
This description, while somewhat precise, forgoes---as it is common---the initialization of the variables (here the starting value of the walk: {\tt 0}), and the technical details about which implementation of the RNG is used.\\

\noindent \begin{minipage}[c]{\linewidth}
\begin{code}{\parbox{.8\textwidth}{\textbf{\textsc{Listing 6:}} Replicated random walk (R$^5$)}\parbox{.161\textwidth}{\hfill \href{https://raw.githubusercontent.com/rougier/random-walk/frontiers/random-walk-R5.py}{raw code}, \href{https://doi.org/10.5281/zenodo.848217}{archive}}}
# Copyright (c) 2017 Nicolas P. Rougier and Fabien C.Y. Benureau
# Release under the BSD 2-clause license
# Tested with CPython 3.6.2 / NumPy 1.12.0 / macOS 10.12.6 / 64 bits architecture
import random
import numpy as np

def _rng(seed):
    """ Return a numpy random number generator initialized with seed
        as it would be with python random generator.
    """
    rng = random.Random()
    rng.seed(seed)
    _, keys, _ = rng.getstate()
    rng = np.random.RandomState()
    state = rng.get_state()
    rng.set_state((state[0], keys[:-1], state[2], state[3], state[4]))
    return rng

def walk(n, seed):
    """ Random walk for n steps """

    rng = _rng(seed)
    steps = 2*(rng.uniform(-1,+1,n) > 0) - 1
    return steps.cumsum().tolist()

if __name__ == '__main__':
    # Unit test
    assert walk(n=10, seed=42) == [1, 0, -1, -2, -1, 0, 1, 0, -1, -2]

    # Random walk for 10 steps, initialization with seed=1
    seed = 1
    path = walk(n=10, seed=seed)

    # Save & display results
    results = {'data': path, 'seed': seed}
    with open("results-R5.txt", "w") as fd:
        fd.write(str(results))
    print(path)
\end{code}
\end{minipage}

It may look innocuous. After all, the
\href{https://docs.python.org/3.6/library/random.html}{Python documentation},
states that \enquote{Python uses the Mersenne Twister as the core generator. It produces
53-bit precision floats and has a period of 2**19937-1}. Someone trying to replicate the work however might choose to use the RNG from the \href{http://www.numpy.org/}{NumPy library}.  The NumPy library is extensively used in the science community, and it provides an implementation of the Mersene Twister generator too. Unfortunately, the way the seed is interpreted by the two implementations is different, yielding different random sequences.\\

Here we are able to replicate exactly\footnote{Striving, as we do here, for a perfect quantitative match may seem unnecessary. Yet, in replication projects, quantitative comparisons are a simple and effective way to verify that the behavior has been reproduced. Moreover, they are particularly helpful to track exactly where the code of a tentative replication fails to reproduce the published results.} the behavior of the pure-Python random walker by setting the internal state of the NumPy RNG appropriately, but only because we have access to specific technical details (the use of the {\tt random} module of the standard Python library of CPython 3.6.1), or to the code itself.\\

But there are still more subtle problems with the description given above.
If we look more closely at it, we can realize
that nothing is said about the specific case of {\tt 0} when generating a step.
Do we have to consider {\tt 0} to be a positive or a negative step? Without
further information and without the original code, it is up to the reader to
decide. Likewise, the description is ambiguous regarding the first element of the walk. Is the initialization value included (it was not in our codes so far)? This slight difference might affect the statistics of short runs.\\

All these ambiguities in the description of an algorithm pile up; some are inconsequential (the 0 case has null probability), but some may affect the results in important ways. They are mostly inconspicuous to the reader and oftentimes, to the writer as well. In fact, the best way to ferret out the ambiguities, big and small, of an article is to replicate it. This is one of the reasons why the ReScience journal \citep{Rougier:2017} was created (the second author, Nicolas Rougier, is one of the editor-in-chief of ReScience). This journal, run entirely by volunteers on GitHub, {\em targets computational research and encourages the explicit replication of already published research, promoting new and open-source implementations in order to ensure that the original research is reproducible}.\\

Code is an integral part of any submission to the ReScience journal. During the review process, reviewers run the submitted code, may criticise its quality and its ease-of-use, and verify the reproduciblity of the replication. The Journal of Open Source Software \citep{Smith:2017} functions similarly: testing the code is a fundamental part of the review process.

\section*{Conclusion}

Throughout the evolution of a small random walk example implemented in Python, we illustrated some of the issues that may plague scientific code. The code may be correct and of good quality, but still many problems may reduce its contribution to scientific knowledge.
To make these problems explicit, we articulated five characteristics that a code should possess to be a useful part of a scientific publication: it should be re-runnable, repeatable, reproducible, reusable and replicable.\\

Running old code on tomorrow’s computer and software stacks may not be possible. But recreating the old code’s execution environment may be: to ensure that the long-term re-runnability of a code, its execution environment must be documented. For our example, a single comment went a long way to transform the R$^0$ code into the R$^1$ (re-runnable) one.\\

Science is built on verifying the results of others. This is harder to do if each execution of the code produce a different result. While for complex parallel workflow this may not be possible, in all instances where it is feasible the code should be repeatable. This allows future researchers to examine exactly how a specific result was produced. Most of the time, what is needed is to set or record the initial state of the pseudo-random number generator, as what done in the R$^2$ (repeatable) version.\\

Even more care is needed to make a code reproducible. The exact execution environment, code and parameters used must be recorded and embedded in the results files, as the R$^3$ (reproducible) version does. Furthermore, the code must be made available as supplementary data with the whole computational workflow, from preprocessing steps to plotting scripts.\\

Making code reusable is a stretch goal that can yield tremendous benefits for you, your team and other researchers. Taken into account during development rather than as an afterthought, simple measures can avoid hours of head-scratching for others, and for yourself—in a few years. Documentation is paramount here, even if it is a single comment per function, as it was done in the R$^4$ (reusable) version.\\

Finally, there is the belief that an article should suffice by itself: the descriptions of the algorithms present in the paper should suffice to re-obtain (to replicate) the published results. For well-written papers that precisely dissociate conceptually significant aspects from irrelevant implementation details, that may be. But scientific practice should not assume the best of cases. Science assumes that errors can crop up everywhere. Every paper is a mistake or a forgotten parameter away from irreproducibility. Replication efforts use the paper first, and then the reproducible code that comes along with it whenever the paper falls short of being precise enough to be reimplemented.\\

In conclusion, the R$^3$ (reproducible) form should be accepted as the minimum
scientific standard \citep{Wilson:2017}. This means this should be actually
checked by reviewers and publishers when code is part of a work worth being
published. But it's hardly the case today.\\


Compared to psychology or biology, the replication issues of computational works have reasonable and efficient solutions. But making sure that these solutions are adopted will not be solved by articles such as this one. Just like in other fields, we have to modify the incentives for the researchers to publish by adopting exigences, enforced domain-wide, on what constitutes an acceptable scientific computational work.


\renewcommand*{\bibfont}{\small}
\printbibliography[title=References]



\end{document}